\documentclass[aps,showpacs,twocolumn,floatfix]{revtex4}
\usepackage{graphicx}

\begin{document}
                                                                                
\title{Quantized electron transport by interference-induced quantum dots of two cross-
travelling surface acoustic waves}
\author{Xiang-Song Chen\footnote{Electronic address: cxs@scu.edu.cn}}
\affiliation{Department of Physics, Sichuan University, Chengdu 610064, China}
\date{October, 2006}
                                                                                
\begin{abstract}
In traditional approaches of obtaining quantized acoustoelectric current, a narrow
channel is fabricated to form quantum dots, which hold a fixed number of electrons 
at a certain depth. We propose a natural way of forming quantum dots without the 
narrow channel, by the interference of two surface acoustic waves (SAWs) propagating 
across each other. A wide transportation area is defined by the usual (but widely 
separated) split-gate structure with another independent gate in between. This design 
can increase the quantized current by one to two orders of magnitude. The three-gate 
structure also allows separate control of the barrier height and the side-gate pinch-off 
voltage, thus avoids current leakage through the area beneath the side gates. 
\end{abstract}

\maketitle

Surface acoustic wave (SAW) travelling in a piezoelectric substrate produces an
accompanying wave of electrostatic potential. When interacting with a two dimensional 
electron gas (2DEG), the wave can drag electrons and drive a current. If a narrow 
channel is put along the SAW direction, it combines with the moving potential trough 
to make a quantum dot, which at a certain depth carries a fixed number of electrons 
due to Coulomb blockade. Devoices based on the above idea have produced quantized 
acoustoelectric current $I=nef$ (where $n$ is the number of electrons in each dot, $f
$ is the SAW frequency) with $f$ in the gigahertz band \cite{Shilton96,Talyanskii97}. 
Such quantized acoustoelectric current can potentially serve as a current standard in 
the nanoamps range \cite{Janssen00}. To increase the current further, Ebbecke {\it et 
al} invented a double-channel structure to duplicate the current \cite{Ebbecke00}. In 
this Letter, we propose a natural way of realizing multiple (up to 10 or 100) 
transportation lines in a single 2DEG, without fabricating physical channels. 

\begin{figure}
\includegraphics[width=8cm]{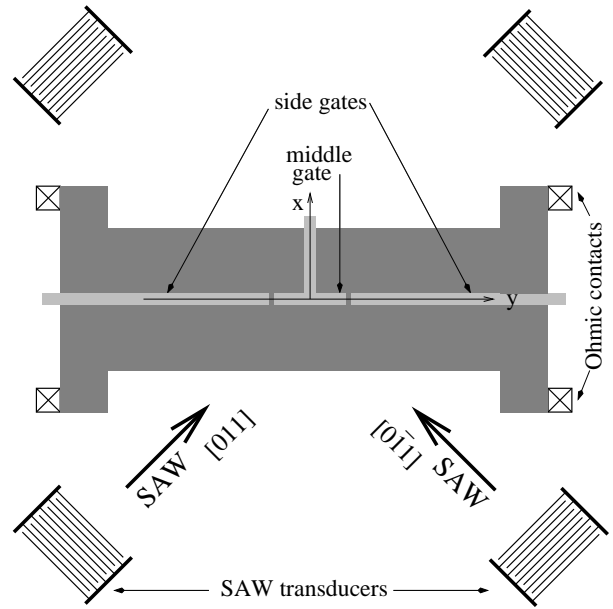}
\caption{Schematic diagram of driving acoustoelectric current by two interfering SAWs. 
A coordinate system is set up as displayed, with the $x$ axis in the $[001]$ 
direction, the $y$ axis in the $[010]$ direction, and the $z$ axis pointing into the 
medium. The two SAWs are travelling in the $[011]$ and $[0\bar 1 1]$ directions, 
respectively.}
\end{figure}

The idea is to make use of the interference pattern of two waves. As depicted in Fig. 
1, two SAWs of the same frequency and perpendicular directions are launched along 
the surface of a piezoelectric substrate such as GaAs/AlGaAs heterostructure. The 
surface is in the $(100)$ plane of the GaAs cubic crystal. The two SAWs are 
travelling in the two perpendicular and symmetric directions $[011]$ and $[0\bar 1 1]
$, respectively. A 2DEG resides some tens of nanometers below the surface. In the 
interfering region of the two SAWs, arrays of potential peaks and wells are formed 
(see Fig. 2). At sufficiently high SAW power, the bottoms of the wells can be 
regarded as quantum dots. The travelling direction of the quantum dots is along the 
middle line of the angle made by the wave vectors of the two SAWs. In our case, it 
is the $[001]$ direction along the $x$ axis. The electron transportation channel is 
defined by the usual split-gate structure, but with the two gates widely separated so 
that a row of many quantum dots can pass together. Namely, it is now a multi-line 
channel. To control the electron number in each dot, a potential barrier must be build 
into the channel as in previous approaches. However, simply applying a voltage to 
the two side gates would not work well, because the channel is now wide. In order 
for the quantum dots to climb over the same potential barrier, a third (middle) gate is 
added into the channel, with its ends very close to the two side gates. The potential 
shape of this configuration is roughly illustrated in Fig. 3 \cite{Davies95}. 

\clearpage

\begin{figure}
\includegraphics[width=6cm]{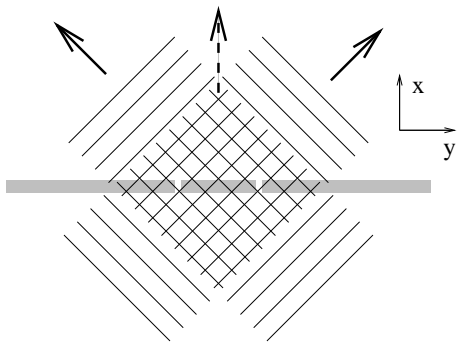}
\caption{The interference region. Shaded squares represent the gates. Parallel lines 
represent the bottoms of the SAW troughs. The cross points are the interference-
induced quantum dots. Solid and dashed arrows indicate the travelling directions of the 
SAWs and quantum dots, respectively. }
\end{figure}

\begin{figure}
\includegraphics[width=8cm]{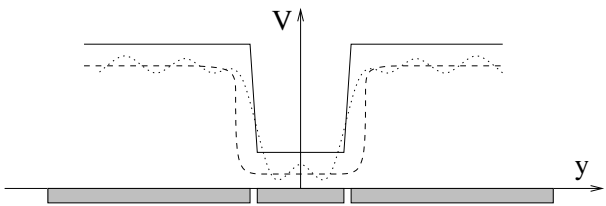}
\caption{A rough illustration of the potential shape in the $y$-$z$ plane. Shaded 
areas indicate where the gates locate. Solid line is the bare potential (without SAW) at 
the surface; dashed and dotted lines are potential at the 2DEG without and with the 
SAW, respectively.}
\end{figure}

If $\eta$ is the average number of transportation lines, the quantized acoustoelectric 
current would be $\eta\times nef$. Since $\eta\sim D/\sqrt 2\lambda$ (where $D$ is 
the channel width, and $\lambda$ is the SAW wavelength) can be made quite large, 
such a device can potentially deliver a quantized current in the sub-microamps range. 
The limitation of this approach is essentially set by the Coulomb interaction between 
the electrons in different dots, which may raise the electron energy too high that the 
dots can bound no electrons. 

The design we proposed brings two extra advantages: First, before encountering the 
gate barrier, the quantum dots have travelled quite some distance, thus have enough 
time to capture electrons and to arrange them into stable bound states. Second, the 
three-gate structure allows separate control of the pinch-off voltage for the side area 
and the barrier height in the transportation region, thus avoids current leakage through 
the area beneath the side gates when the barrier height is decreased. We encourage 
experimental investigation of the interaction of two interfering SAWs with the 2DEG, 
especially in connection with the theoretical \cite{Aizin98,Gumbs99,Flensberg99,
Galperin01,Robinson01,Gumbs06} and experimental \cite{Cunningham99,Fletcher03,
Gloos04,Robinson05,Kataoka06} efforts towards a thorough understanding of the 
quantization phenomenon of acoustoelectric current.

The author thanks Prof. Jie Gao for inspiring discussions. 
This work was supported in part by the China NSF (grant 10475057).


\begin{thebibliography}{99}
\bibitem{Shilton96} J. M. Shilton, V. L. Talyanskii, M. Pepper, D. A. Ritchie, J. E. F. 
Frost, C. J. B. Foar, C. G. Smith, and G. A. C. Jones, J. Phys.: Condens. Matter {\bf 
38}, L531 (1996). 

\bibitem{Talyanskii97} V. I. Talyanskii, J. M. Shilton, M. Pepper, C. G. Smith, C. J. 
B. Ford, E. H. Linfield, D. A. Ritchie, and G. A. C. Jones, Phys. Rev. B {\bf 56}, 
15180 (1997). 

\bibitem{Janssen00} T. J. M. Janssen and A. Hartland, Physica B {\bf 284-288}, 1790 
(2000). 

\bibitem{Ebbecke00} J. Ebbecke, G. Bastian, M. Bl\"ocke, K. Pierz, and F. J. Ahlers, 
Appl. Phy. Lett. {\bf 77}, 2601 (2000).

\bibitem{Davies95} J. H. Davies, I. A. Larkin, and E. V. Sukhorukov, J. Appl. Phys. 
{\bf 77}, 4504 (1995).


\bibitem{Aizin98} G. R. A\v\i zin, G. Gumbs, and M. Pepper, Phys. Rev. B {\bf 58}, 
10589 (1998).

\bibitem{Gumbs99} G. Gumbs, G. R. A\v\i zin, and M. Pepper, Phys. Rev. B {\bf 
60}, R13954 (1999). 

\bibitem{Flensberg99}, K. Flensberg, Q. Niu, and M. Pustilnik, Phys. Rev. B {\bf 60}, 
R16291 (1999).

\bibitem{Galperin01} Y. M. Galperin, O. Entin-Wholman, and Y. Levinson, Phys. Rev. 
B {\bf 63}, 153309 (2001). 

\bibitem{Robinson01} A. M. Robinson and C. H. W. Barnes, Phys. Rev. B {\bf 63}, 
165418 (2001).

\bibitem{Gumbs06} G. Gumbs and Y. Abranyos, Phys. Rev. B {\bf 73}, 085303 
(2006).  


\bibitem{Cunningham99} J. Cunningham, V. I. Talyanskii, J. M. Shilton, M. Pepper, 
M. Y. Simmons, and D. A. Ritchie, Phys. Rev. B {\bf 60}, 4850 {1999}. 

\bibitem{Fletcher03} N. E. Fletcher, J. Ebbecke, T. J. B. M. Janssen, F. J. Ahlers, M. 
Pepper, H. E. Beere, and D. A. Ritchie, Phys. Rev. B {\bf 68}, 245310 (2003). 

\bibitem{Gloos04} K. Gloos, P. Utko, J. B. Hansen, and P. E. Lindelof, Phys. Rev. B 
{\bf 70}, 235345 (2004). 

\bibitem{Robinson05} A.M. Robinson and V. I. Talyanskii, Phys. Rev. Lett. {\bf 95}, 
247202 (2005). 

\bibitem{Kataoka06} M. Kataoka, C. H. W. Barnes, H. E. Beere, D. A. Ritchie, and 
M. Pepper, Phys. Rev. B {\bf 74}, 085302 (2006). 
\end{thebibliography}
\end{document}